# Decoding a Complex Visualization in a Science Museum – An Empirical Study

Joyce Ma, Kwan-Liu Ma, *Fellow, IEEE*, and Jennifer Frazier

**Abstract**—This study describes a detailed analysis of museum visitors' decoding process as they used a visualization designed to support exploration of a large, complex dataset. Quantitative and qualitative analyses revealed that it took, on average, 43 seconds for visitors to decode enough of the visualization to see patterns and relationships in the underlying data represented, and 54 seconds to arrive at their first correct data interpretation. Furthermore, visitors decoded throughout and not only upon initial use of the visualization. The study analyzed think-aloud data to identify issues visitors had mapping the visual representations to their intended referents, examine why they occurred, and consider if and how these decoding issues were resolved. The paper also describes how multiple visual encodings both helped and hindered decoding and concludes with implications on the design and adaptation of visualizations for informal science learning venues.

✦

## 1 INTRODUCTION

The last decade has seen several large-scale efforts by science museums and other informal learning settings to create visualizations of the large, complex datasets increasingly produced and used by scientists. These visualizations provide informal learners an opportunity to engage with critical new areas of science and foster important data literacy skills. Recent projects include the National Oceanic and Atmospheric Administration's (NOAA's) Science on a Sphere program, which visualizes environmental data on a spherical display [1]; the DeepTree exhibit, which allows visitors to interact with and explore an evolutionary tree of life with over 70,000 species [2]; and the Living Liquid project, which created interactive visualizations of the micro and macroscopic life in the world's oceans [3]. These visualizations must address design challenges unique to museums and other informal learning contexts: the short time visitors spend at an exhibit; visitors who have little prior knowledge of the data or how it was collected; supporting multiple users for the majority of visitors who visit in groups; and, initiating and sustaining interest in the free-choice learning environment where there is no set sequence of exhibits and other competing attractions [4]–[7]. Identifying promising ways to support visitor engagement with visualizations is an ongoing effort in the museum field, which is eager to give the public access to the scientific discoveries made possible with large datasets.

A critical first step in engaging the public with scientific phenomena and content in a visualization is decoding. Decoding is the process by which visitors map the visual elements within a visualization to the data and data relationships that they are meant to represent. Decoding is fundamental to understanding the patterns and relationships among the data variables that characterize the underlying phenomenon and is, therefore, a prerequisite to any data exploration and content understanding made possible with visualizations.

The goal of the work described here is to provide a detailed examination of the process by which museum visitors decode a visualization of a large, complex dataset. More specifically, this case study addresses the questions:

- How do visitors decode a visualization of complex scientific data? By complex, we mean a large dataset with multiple, inter-related variables of different types.
- What aspects of the visualization design help and/or hinder the decoding process for visitors?
- What lessons can be applied to the future design and use of visualizations for the museum context?

The work takes a mixed-methods approach to gain insights into the process by which museum visitors decode a visualization, called *Plankton Populations*, that was developed to give visitors access to a scientific dataset that they could explore. Think-aloud data were collected from visitors recruited to use the visualization on the museum floor to unpack the decoding process and identify supports and impediments to visitors' decoding efforts. Naturalistic observation data of uncued visitors at *Plankton Populations* were used to identify how the decoding process may have impacted visitors' attempts at data exploration. This work seeks to build on and extend the visualization field's understanding of how the public makes sense of visualizations of the complex datasets that increasingly define scientific discovery.

## 2 BACKGROUND

### 2.1 Conceptual Framework

This study takes a cognitive perspective that assumes that decoding a visualization depends on constructing a mental mapping between its visual elements, any supporting text, and the data they are intended to represent (also called referents), as posited in Palmer's theory of external representations [8]. This work does not further define the form of these mental mappings, whether they are schemas, frames, or mental simulations. This cognitive view focuses the work on explicating the internal mental mapping instead of the social, activity or cultural interactions that would characterize a socio-cultural perspective.

### 2.2 Related Work

While the museum field is actively creating visualizations for the public, to our knowledge there has been no published work analyzing the process by which a museum visitor decodes a visualization in an informal learning context. To date, there have been some studies that interviewed visitors after using a visualization to assess their familiarity with and ability to decipher representations. For example, a large cross-museum study on data literacy was conducted by Börner et al. [9] to ascertain which types of data visualizations (i.e.,

- *Joyce Ma and Jennifer Frazier are with the Exploratorium in San Francisco, e-mail: [jma, jfrazier]@exploratorium.edu.*
- *Kwan-Liu Ma is with the University of California, Davis, e-mail: ma@cs.ucdavis.edu.*

charts, maps, graphs, or network layouts) visitors recognized and were able to read. Their study found visitors were familiar with basic maps, charts, and graphs, and that visitors self-reported colors, lines, and text as important for reading visualizations. In addition, exhibit evaluations have documented museum visitors' challenges deciphering color in visualizations ranging from nanoscale structures to ocean satellite imagery; visitors have been found to link colors to temperature despite design intentions [10], [11]. Aside from these studies related to decoding, most published work on visualizations of large scientific datasets have focused on explicating design considerations particular to the museum context [3], [12], looking at patterns of collaboration at these visualizations [13], and describing the nature of learning [2], all of which contributes to our understanding of how visitors use visualizations in settings of informal learning. To our knowledge, this study represents the first detailed analysis of visitors' decoding process as they use a visualization in an informal learning context.

However, there has been extensive work in formal education and learning research on the use of multiple external representations (MERs) in computer-based learning environments [14]–[16] that can give insights into learners' decoding process. MERs bundle several dynamic representations such as animations, graphs, maps, and other interactive visual representations into one program [15]. And, with a complex scientific dataset, a visualization often needs to implement different visual representations to capture different aspects of its richness. In particular, Ainsworth's Design Function Task (DeFT) framework [17], formulated to explain how MERs may work together to support learning, can be helpful in explaining how decoding may be helped or hindered by multiple visual representations used in the same visualization. According to Ainsworth, MERs can serve three important functions, with the first two playing critical roles in decoding. First, MERs can complement each other to provide multiple points of access into the dataset by encoding the same information in different ways. Second, MERs can constrain decoding, with one representation, typically the more familiar one, helping to decipher another. Third, coordinating between the different ways in which data can be represented is a critical part to building a robust understanding of the relationships among different data. Despite the benefits they provide learners, using MERs requires relating, or linking, the different representations, which can increase the cognitive load for a person already struggling to understand a single visual representation [15].

To date, there has been one study that looked at MERs in the museum context. Wang and Yoon's work [18] investigated how three dynamic visualizations (digital augmentation, computer simulation, and animation) worked together to support visitors' understanding of Bernoulli's Principle at a hands-on exhibit. It found evidence of increased learning with multiple visualizations. The three visualizations, however, were not integrated into one, and were all simulations of a physical principle as opposed to rich datasets to be explored by visitors. The data presented in this study extends our understanding of MERs by looking at their use when integrated into one visualization for data exploration at a museum.

Within the Human-Computer Interactions (HCI) sensemaking literature, recent work by Lee et al. [19] looked in detail at how university study subjects made sense of three different types of visualizations (i.e., a parallel-coordinates plot, a chord diagram, and a treemap) previously unfamiliar to them. In examining how frames of visual encoding, the mapping between a visual (or textual) representation and its referent, are constructed, used, and modified, Lee et al. articulated how a non-expert may decode a new visualization in their NOvice's information VIsualization Sensemaking (NOVIS) model. This study, however, differs from the work by Lee et al. in two regards; unlike the simple visualizations considered in formulating NOVIS, *Plankton Populations* includes multiple visual representations for a much more complex dataset. In addition, a museum visualization typically tries to include visual representations (e.g., a map, a timeline) that are somewhat familiar to visitors in order to help ready comprehension. NOVIS, nonetheless, provides a touchstone for the work reported here.

## 3 THE PLANKTON POPULATIONS VISUALIZATION

This case study examined the process by which museum visitors decoded *Plankton Populations*, an interactive visualization of marine microbes, or plankton, in the global ocean. *Plankton Populations* is part of the life sciences exhibit collection at the Exploratorium, a science museum in San Francisco, California, that has over 850,000 visitors a year. The exhibit is one of the few interactive visualizations of an authentic scientific dataset created for the museum environment. The version of the exhibit used in this study went through several rounds of iterative development and formative evaluation, and was subsequently summatively evaluated by Inverness Research Inc., an independent exhibit evaluation group. The summative evaluation found evidence that *Plankton Populations* initiated data exploration and enabled content understanding [20]. *Plankton Populations*, therefore, provided a good case to study museum visitors' decoding processes and challenges they may encounter in using visualizations.

### 3.1 Description

The *Plankton Populations* visualization is an adaptation of the Darwin Project, a supercomputer-based simulation of the distribution of marine microbes that uses environmental and biological data from satellite, buoy, environmental sampling, and laboratory studies. The *Plankton Populations* visualization is on a large, 55-inch, multi-touch table-top interactive screen, surrounded by a backlit static label containing text, images, and a legend (Fig 1). Although anyone can approach and use it, *Plankton Populations*' target audience is visitors eight years old or older, who should have a basic understanding of ecosystems necessary to interpret the visualizations, including the idea that living things are dependent on their environment [21].

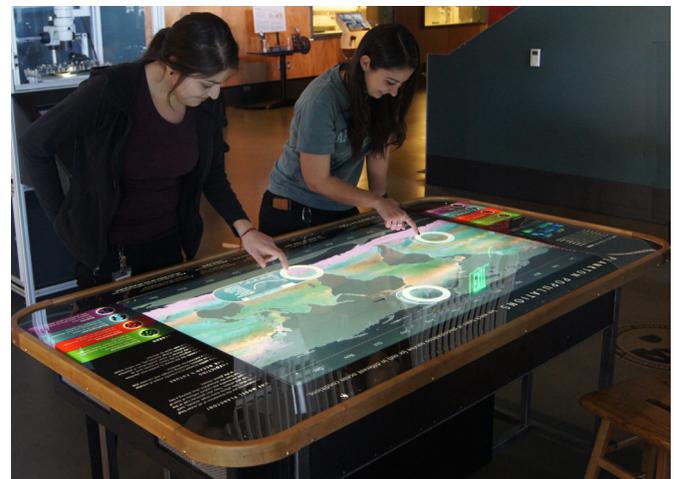

Fig. 1. Users at the *Plankton Populations* exhibit in the Living Systems Gallery of the Exploratorium.

*Plankton Populations* is an updated version of an exhibit previously called Living Liquid, whose design and development was described in depth in a prior publication [3], which focused on visualization design and not on analyzing the decoding process. To summarize, the primary visualization is an animation of the areas where four different types of plankton live in the ocean, represented by four colors. The animation shows how plankton distribution changes through the course of the year, with the area and location of the colors changing over time. The passage of time is also indicated by a monthly timeline running along the top and bottom of the map. Overlaid on the global ocean are three interactive "lenses" visitors can move around to explore different ocean regions. When a lens is placed at a location in the ocean for half a second or more, the portion of the global map under the lens fades while icons representing the relative number of different types of plankton at that location appear. (See Fig 2 and supplemental materials.)

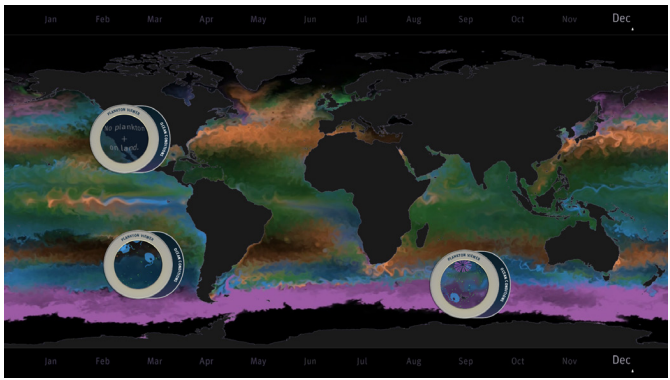

Fig. 2. Screenshot of the *Plankton Populations* visualization.

The updated version of the exhibit gives visitors the ability to examine the environmental data at the location of a lens. A tab on the side of the lens labelled 'Ocean Conditions' allows visitors to access a call-out with graphs of the three key environmental variables: light, nitrogen, and silica (Fig 3A) that correlate with the types and quantities of plankton found. When a tab is selected for an environmental variable, a graph of the amount of that variable found in that location over the course of the year is shown (Fig 3B). This graph is dynamic and changes if the lens is moved. Providing these environmental data allows visitors to ask and answer questions about why different types and amounts of plankton are growing in a location.

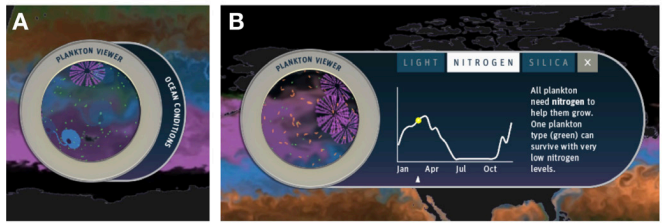

Fig. 3. (A) Close up of the interactive "lens" showing plankton number and type, with call-out tab closed; (B) close up of the lens with call-out tab open to show the levels of environmental variables.

To make room for the ocean conditions graphs, the legend describing the four plankton types was moved to the static label that surrounds the screen (Fig 4). In prior versions, this legend had been accessible in the call-out tab that is now used for showing environmental variables.

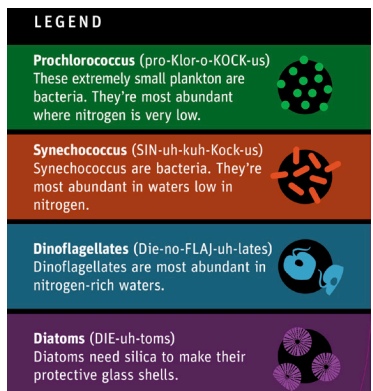

Fig. 4. Legend from the label surrounding the screen, with plankton encoding.

## 3.2 Design Trade-offs

Creating *Plankton Populations,* a visualization to promote exploration of a complex dataset by non-experts, required a delicate balance of providing easy entry points to initiate exploration while also providing access to the richness of the dataset to sustain engagement. This is a challenge posed in the design of any exhibit trying to foster active, prolonged engagement [22]. In this case, design entailed making difficult trade-offs that could impact decoding. The following describes those design choices, their rationale, and possible decoding issues they raised, which were then examined in this study.

**Prioritizing one aspect of the dataset over another**. *Plankton Populations* encodes a complex dataset with multiple variables and their relationships in a dynamic visualization. Prior empirical and theoretical work highlights the need to manage the high cognitive load that can result with animations [23]–[25]. In an attempt to control the visual complexity, the design emphasizes the location of different plankton types while initially downplaying the environmental variables. That is, the geographic distribution of plankton type is immediately visible when visitors first step up to the exhibit (1) in the color regions in the global ocean and (2) in the types of plankton shown in lenses. Alternatively, data for the environmental variables (i.e., the silica, light, and nitrogen levels that change with location and time) are only accessible upon demand, appearing when visitors tap the side of a lens. The intention is to encourage visitors to first spend time using the lens to explore the geographic distribution of plankton in different parts of the ocean and only then begin to correlate plankton to changing environmental conditions. A detailed look at visitors' decoding process provided the opportunity to address the question: Was layering secondary data access a successful strategy to help ease decoding?

**Using color to attract and support engagement.** As an exhibit in a free-choice learning environment, *Plankton Populations* needs to attract as well as hold visitors' attention amidst the distractions and competing stimuli found on an active museum floor [26]. To design a visually attractive exhibit, we chose to use color to encode the four different types of plankton in the dataset. An earlier study [3] found that using colors to represent plankton types prompted visitors to ask questions about the geographic distribution of plankton, the part of the data we hope visitors would explore first. The color encoding, therefore, also serves to give visitors immediate access to the data. However, given the problems prior work [10], [11] and early *Plankton Populations* formative evaluations surfaced on using color in museum data visualizations, the team did look to the visualization literature and consulted with several visualization experts from computer science and data arts to find alternate encodings. While several alternatives were considered (e.g., bar charts or dots to represent plankton amount and type), no better options were found. By taking a closer look at this encoding choice, this study allowed us to address the questions: Did color help or hinder immediate access to the plankton data? More importantly, why might visitors find the color encoding challenging to decipher?

**Using multiple visual representations to constrain and complement decoding**. Guided by Ainsworth's DeFT framework [17], *Plankton Populations* uses MERs to complement and constrain one another. For example, the spatial distribution of plankton is encoded two ways: with four colors on a global map that denotes the four different plankton types, and with color-coded plankton icons that appear inside a lens, encoding the plankton found in that location. In this case, these two representations complement each other and allow two different points of access to plankton type.

Alternatively, other MERs were designed to constrain, or support, the decoding of each other. For example, the 'No Plankton On Land' text that appears inside the lens when placed over a landmass is meant to help visitors decipher the map as well as the icons that appear in the lens, while a legend on the exhibit's static label maps the icons and colors to each other and to their shared referent, the different plankton types. Where possible, *Plankton Populations* incorporates familiar visual representations. More specifically, *Plankton Populations* uses a map, one of the most commonly recognized visualizations among museum visitors [9], to encode location data. And, simplified icons of microorganisms are used instead of micrographs to support recognition [27]. The design, thereby, aims to use more familiar visual representations, which are more readily deciphered [9], to help constrain the decoding of the less familiar.

We note that whether MERs serve complementary or constraining functions is fluid, changing with context. The lens icons, as a complement to the map colors, could provide the primary means for one visitor to explore plankton distribution in the visualization. For another visitor, these same icons could be used to determine which map colors corresponds to which plankton type.

Prior work on MERs in formal education points to the advantages of using MERs but also cautions designers for the need to link MERs [15], [28]. *Plankton Populations* does this in multiple ways. When a lens, designed to connote a magnifying glass is placed at a location for one second or more, the map under the lens fades and icons representing the different plankton types appear inside the lens, mimicking the focusing of a microscope. Color is also used to link MERs, with the same colors being used for overall regions of the ocean where a plankton is dominant, for that plankton icon inside the lenses, and in the legend linking the two. Close proximity linking, however, was not always possible. For example, the legend mapping the colors to the four plankton types is placed in the static label on the side of the *Plankton Populations* exhibit to make room for the graphs of the ocean conditions that need to change with the changing location of the lenses. Using MERs adds to the visual complexity of the visualization. In this study, we looked at the questions: Was there any evidence that MERs helped decoding, and how was that manifested?

## 4 METHOD

We used a mixed method design to study visitors' decoding processes at *Plankton Populations*. The study collected two sets of data: (1) think-aloud data from cued visitors, who were recruited and consented to participate in a study; and (2) naturalistic observation data of a second set of visitors' behavior at the standalone, un-facilitated exhibit. The majority of the analysis was conducted with the think-aloud data, which provided the verbalization of visitors' thinking necessary for detailed qualitative analysis. The observation data were used primarily as a supplement to the think-aloud data as a check on participant reactivity or any pleasing bias from visitors recruited by staff.

### 4.1 Think-Aloud Data

For the think-aloud study, we approached every third visitor who crossed a predetermined imaginary line near *Plankton Populations* and who appeared to meet our selection guidelines, which were that the visitor: (1) was at least eight years old, (2) was with one other person, (3) spoke English, and (4) was near an accompanying adult who could give informed consent if the selected visitor was a minor. We asked this person if s/he and one other person in their visiting group would be willing to participate in a study to help in the development of a new exhibit at the Exploratorium, and to be videotaped in the process. We chose to recruit dyads because we believed that visitors would have an easier time verbalizing and sharing their thoughts with a person they came with rather than a staff member. However, a researcher was there to prompt visitors to talk in case they fell silent. Upon consent, participants were asked to think aloud, that is, talk about what they were thinking and trying to do as they used the exhibit, and to answer a few questions immediately afterward. In total, we recorded 56 dyads' conversations and interactions at *Plankton Populations* over twelve days of data collection. The demographic information of the think-aloud participants is given in Table 1.

Table 1. Think aloud participants' demographic information.

| Gender | Count | Age | Count |
| --- | --- | --- | --- |
| Female - Female | 16 | Adults | 42 |
| Female - Male | 34 | Adult-Minor | 8 |
| Male - Male | 6 | Minors | 6 |

We reviewed the think-aloud recordings using Datavyu [29] software (Fig 5), listening for two categories of talk: (1) decoding comments and (2) data interpretations. A decoding comment is a remark about a visual encoding used in *Plankton Populations*, as listed in Table 2. Since decoding comments speak to the visualization's design attributes, an analysis of these comments can help refine our understanding of the applicability of those design decisions in the museum context. In this analysis, decoding comments encompass visitors simply noting an encoding, asking questions about it, or mapping the visual encoding to its referent, which is the underlying data and concept it represents. For each such remark, we identified the visual encoding, the data variable if the dyad attempted to map the encoding to its referent, and whether or not the mapping was 'correct.' A 'correct' mapping is narrowly defined in this coding scheme as one that was intended by the visualization designers, even though an 'incorrect' mapping may in fact be a legitimate way of decoding the visualization. Table 3 provides examples of decoding comments with their applied codes.

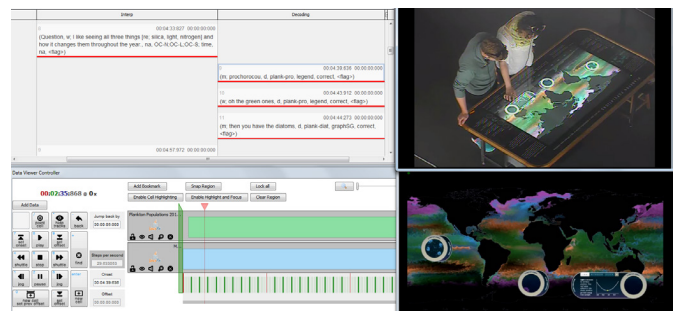

Fig. 5. Screenshot of Datavyu video coding. Videos are right, coding entry is upper left, and play head control is lower left.

Table 2. Visual encodings in *Plankton Populations*.

| Visual Encoding | Data Variable (Referent) |
| --- | --- |
| Map | location |
| Timeline | time |
| Lens Icons | plankton |
| Graph | environment |
| Color | plankton |

Table 3. Coding scheme for decoding comments.

| Code | Description |
| --- | --- |
| Decode | A remark about a visual encoding |
| - Encoding | The visual encoding mentioned |
| - Referent | The referent data variable if the decoding statement maps a visual encoding to its meaning |
| - Correct? | Indicates if the mapping was intended by the visualization designers |
| **Examples** | |
| (Points to nitrogen graph.) This is not a very high nitrogen level. | Encoding: graph<br>Referent: environment<br>Correct?: correct |
| Oh, I see some currents over here. | Encoding: color<br>Referent: current<br>Correct?: incorrect |
| I like the shapes, seeing the shapes and the colors [inside the lens]. | Encoding: lens<br>Referent: n/a<br>Correct?: n/a |

Data interpretations are remarks that describe a pattern or relationship among the different data variables in the *Plankton Populations* dataset, listed in Table 4. They are indications that visitors are engaging with the scientific content embodied in the data. For each data interpretation, we noted the data variables and the visual encodings used to note that relationship or pattern. When it was obvious that the data interpretation was based on an incorrect decoding, we noted it as such. Table 5 summarizes the data interpretation code and provides examples of its application to visitor talk.

Table 4. Key relationships between data variables in the dataset that visitors can explore.

| Data Relationships | Description |
|---|---|
| plankton-place | Content: Different plankton types live in different areas. |
| plankton-time | Content: Plankton populations change over time. |
| env-place | Content: Environmental conditions vary by location. |
| env-time | Content: Environmental conditions change over time. |
| plankton-env | Content: Environmental conditions determine the distribution of different plankton types. |

Table 5. Coding scheme for data interpretation talk.

| Code | Description |
|---|---|
| Data Interpretation | A remark about a pattern or relationship among the data variables in the dataset |
| - Variables | The data variables in the pattern |
| - Encoding | The visual encoding used in the interpretation |
| - Correct? | Marked as incorrect if the data interpretation is based on an erroneous decoding |
| **Examples** | |
| [There's] plenty of green ones there (looking at lens on map), and it's nitrogen rich (with nitrogen graph open). | Variables: place-plankton-env  Encoding: lens-map-graph  Correct?: correct |
| So, I understand in summer the poles are dead. | Variables: place-plankton-time  Encoding: lens-map-timeline  Correct?: correct |
| (Looking near Australia with lens) there's everything [every type of plankton] | Variables: place-plankton  Encoding: lens-map  Correct?: correct |

The unit of analysis was the dyad and not the individual. Although we noted who within the pair made each comment, none of the subsequent analysis distinguished between the two visitors. This decision was made because visitors often completed each other's sentences, making it difficult to attribute an articulated thought to just the speaker.

We defined this coding scheme by first listening for decoding and interpretation talk in the think-aloud recordings from three randomly selected dyads to determine how these comments manifest. A pilot coding scheme was defined based on this first listen, and then two data coders were asked to work independently to apply the codes to each of the three selected recordings. After each think-aloud recording, the two coders discussed how they applied the codes and worked together to resolve any coding discrepancies. With each successive iteration, the coders revised and refined the coding scheme.

The remaining videos were assigned to two coders in four separate batches. Each coder independently applied the coding scheme to each dyad's recorded interactions and talk at *Plankton Populations*, one batch at a time. When the two data coders finished a batch, we calculated the interrater reliability statistic, informed the coders which recordings had been selected for interrater assessment in that batch, and asked them to discuss and resolve any coding discrepancies in the shared recordings. This helped the coders realign with each other and with the coding scheme throughout the qualitative coding effort. Fourteen out of the total 56 dyads (25%) were randomly selected and assigned to both coders to assess interrater reliability. The Cohen's Kappa statistic, a conservative measure that corrects for chance agreement, was 0.76 and 0.75 for detecting decoding and data interpretation talk in visitors' think-aloud recordings, respectively, indicating substantial agreement according to Landis and Koch [30].

### 4.2 Naturalistic Observation Data

We recorded approximately 28 hours of observation data with ceiling cameras over six weekend days. The overhead cameras allowed us to unobtrusively capture visitors' behavior while protecting their anonymity (Fig 6). A data coder watched the video and systematically selected every third visitor who approached and stopped at *Plankton Populations* for more than five seconds, noting their apparent age group (i.e., under eight years old, minor, or adult) and the total time spent at the exhibit. We then eliminated from the data corpus any visitor who looked under eight, and therefore outside of this exhibit's target age group, or who spent less than five seconds at the exhibit, an insufficient time for even the most cursory interaction. Our naturalistic observation totalled 160 visitors.

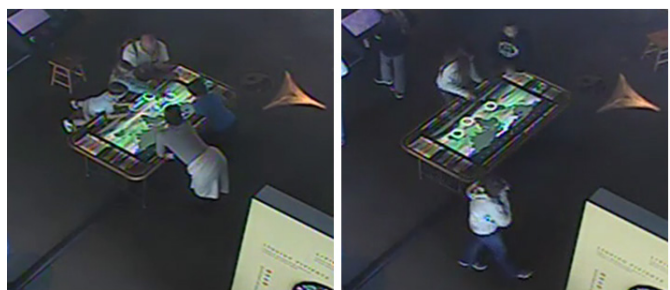

Fig. 6. Screenshots of low resolution overhead video collected for naturalistic observation.

## 5 RESULTS AND DISCUSSION

The goal of this analysis was to unpack the changes in visitors' thinking, at a fine temporal grain, as they attempted to decode the visualization while exploring the underlying data content. This study was focused on understanding decoding to inform future work, not to assess the knowledge or skills acquired by visitors nor to conduct an outcome assessment of the exhibit, which was summatively evaluated separately.

### 5.1 Visitors Decoded Throughout their Interactions

To help understand how decoding happened, we plotted visitors' comments over time. The resulting graph (Fig 7) reveals several patterns. First, the presence of data interpretation remarks indicates that visitors were using the visualization to examine the data (e.g., to see how plankton changes with location and time). Of the 56 dyads, only one group failed to make any data interpretations during their think-aloud[1].

Second, there was a preponderance of decoding talk, especially before the first data interpretation. This was an expected finding since deciphering the visual representations is necessary to making connections and seeing patterns in the data. However, decoding was not confined to the period before the first data interpretation. Instead, decoding statements continued well into each dyad's think-aloud, interweaving with data interpretations. We would expect this type of pattern for a complex dataset with multiple visual elements used to encode different aspects of the larger data. It would have been surprising to find visitors systematically learning all the encodings before engaging in data exploration. We note that Lee et al. [19] found a similar pattern even with relatively simpler visualizations.

---

[1] This dyad of two children said little despite repeated attempts from the data collector encouraging them to talk about what they were thinking and doing.

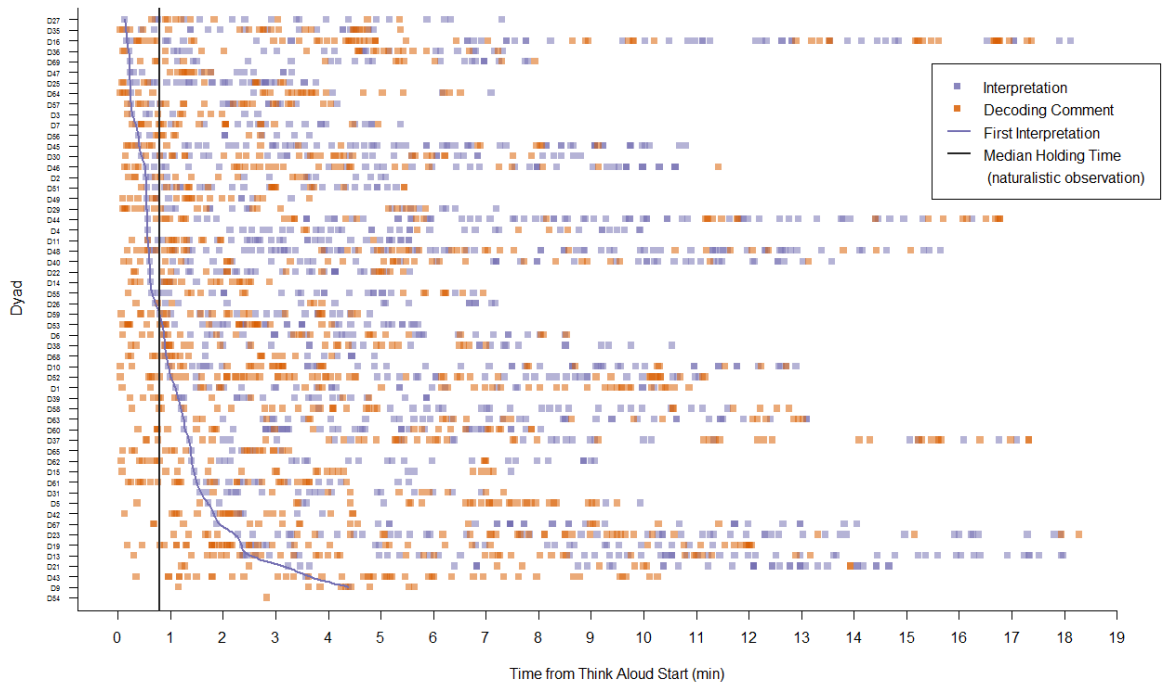

Fig. 7. Timeline of decoding and data interpretation comments in study participants' think-alouds. More saturated squares indicate multiple statements occurring close to one another.

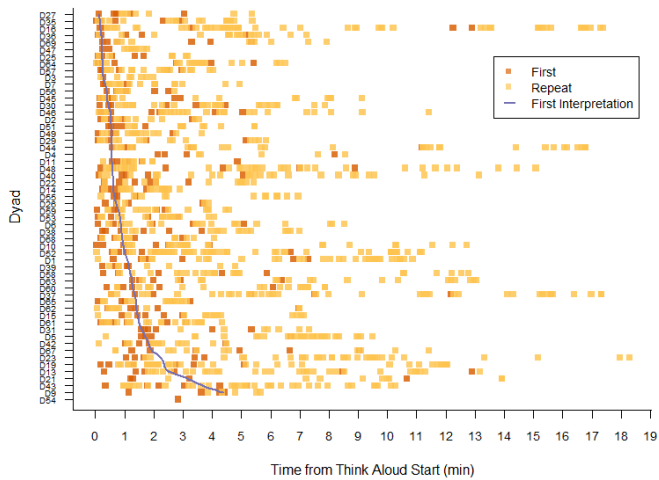

Fig. 8. Timeline of decoding comments indicating when an encoding is mentioned for the first time.

A closer look at the decoding statements (Fig 8) revealed that 73%[2] of each dyad's decoding talk was regarding a visual encoding that they had already mentioned. For example, Dyad11 talked about the lens icons for the first time 34 seconds into their think-aloud and then again about 1 minute in as they noticed how the icons changed with the data they represented. The ongoing decoding talk was typically not of visitors deciphering subsequent, different encodings but rather revisiting a prior encoding. This suggests that visitors did not 'learn' a visual encoding just once. Instead, decoding a visual element was an ongoing act of construction. In fact, to visitors a visual encoding could appear as new when the data it represents change significantly.

---

[2] Mean across the 56 dyads.

Transcript excerpt from Dyad11
0:34   That's cool how it changes, the different creatures [shown in the lens].
0:58   What kind of plankton are these? [shown in the lens]
1:02   Synechococcus

**Design Implication**. This result suggests that visitors need support in decoding throughout their exhibit interaction, not just upon initial engagement. This support may be especially critical in dynamic visualizations, in which the underlying data may be changing.

### 5.2   Time to First (Correct) Interpretation was Long for the Museum Context

We looked at the time it took for the dyads to make their first data interpretation and found that the Time to their First Interpretation (TFI) was 43 seconds, median, while the Time to their First *Correct* Interpretation (TFCI) was 54 seconds, median. Considering the museum context, where the total holding time at an exhibit is measured in seconds, 43 seconds is a long time for a visitor to arrive at his/her first data interpretation. As a point of reference, the holding times for exhibits in an earlier Exploratorium life sciences collection ranged from 12 to 149 seconds [31].

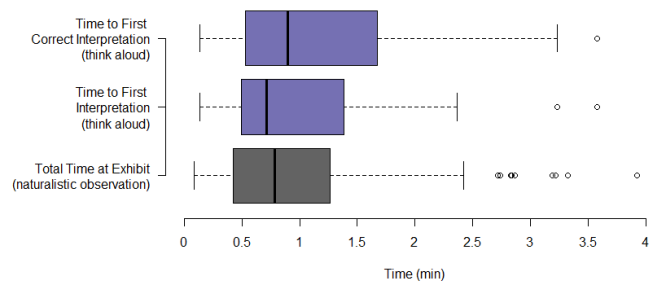

Fig. 9. Boxplot comparing the time the think-aloud participants took to arrive at their first (correct) data interpretation and the total time uncued visitors stayed at Plankton Populations in naturalistic observations.

Looking at the naturalistic observation data, we found that the median of the *total time* uncued visitors stayed at *Plankton Populations* was 47 seconds, only a few seconds more than the median TFI and a few seconds less than their median TFCI as shown in Fig 9. That is, likely half of the visitors behaving naturally at the exhibit were not engaging in data exploration, instead spending their time trying to decode the visual encodings. This suggests that as a standalone, *Plankton Populations* likely does not initiate data exploration with a subset of users.

**Design Implication**. Given the short time visitors spend at an exhibit, a museum visualization needs to help visitors quickly decode enough of the visualization to allow them to start data exploration.

### 5.3 Including Secondary Data Delayed Interpretation

To elucidate how we may lower barriers to data exploration, we revisited the trade-offs we made in designing *Plankton Populations* that may have helped or hindered decoding. One way by which *Plankton Populations* tries to manage complexity is to distribute the information encompassed in a complex scientific dataset across different but related visual encodings. Different aspects of the data can then be introduced after the more fundamental relationships are explored. To do so, *Plankton Populations* prioritizes visualizing the place-plankton relationship with colors and lenses over the relationships with the environmental variables, which only appear in graphs accessible by tapping on the side of a lens.

However, it can be challenging to structure visitors' interactions in the museum context [5], [7]. For example, a visitor can come upon the exhibit in any state, with the more detailed views of environmental conditions already open. Also, visitors may not approach data exploration at a museum exhibit with a systematic bent and may opt to open all available windows and easily become overwhelmed. This becomes more challenging with a multiuser table where multiple visitors may be at different stages of exploration [32].

Looking at visitors' first *correct* data interpretation, we found that a majority of these comments was about the location of the plankton (29/55). The second most frequent first correct interpretation was the correlation between environmental conditions and plankton (19/55), but all of these 19 visitors read this in the label text instead of deciphering and coordinating the visual encodings. This finding gives some limited support that visitors were initially focused on the plankton's spatial distribution in *Plankton Populations*.

But, did the visual encoding of the environmental variables delay visitors' data interpretation? To shed light on this question, we compared dyads who did and did not mention graphs in any way before their first correct data interpretation comment. We used a Box-Cox transformation, with $\lambda$ = 0.023, to correct the right-skewed TFCI data and conducted a t-test on the resulting statistically normal dataset. The comparison found that dyads who noted the graphs ($n$ = 29) took longer (reverse-transformed $M$ = 68 seconds) than those who did not ($n$ = 26, reverse-transformed $M$ = 38 seconds); $t(50)$ = 2.53, $p$ = 0.015). Absent a control group, we could not attribute the increase TFCI to the graphs. But, this finding suggests that the visual elements (i.e., the graphs) encoding the secondary data variables were not completely ignored on initial interaction, and that when noticed by visitors, they increased the TFCI.

**Design Implication**. Visual encodings of secondary data should be added judiciously since they may easily overwhelm visitors. If secondary data are added, designs should include strategies such as timed closing of pop-out windows, small personalized views, or anchored stations.

### 5.4 Decoding Color was Challenging for Multiple Reasons

Originally, the color encoding was intended to not only provide visual appeal but also immediate access to the spatial distribution of plankton populations. We were, therefore, surprised that of the 29 dyads whose first data interpretation was of the plankton's spatial distributions, none relied on the color encoding. Instead, all 29 used the lenses, which gave the more local view of the plankton found at the lenses' positions on the map. Furthermore, when we looked at all 56 visitors, a large majority (44/56) mentioned color but the median time to its first mention occurred 02:54 into their think-alouds. Although Lee et al. [19] also found that novices did not always verbalize their sensemaking when they first encountered a visualization, it was, nonetheless, surprising to find such a dearth of color talk at the beginning of visitors' exhibit experiences. There was, however, one dyad, Dyad 16, who spent the majority of their time systematically investigating each variable to understand what the colors mapped to (see supplemental material).

The color encoding was, therefore, ineffectual in giving visitors ready access to the data for exploration. Instead, the lenses were the entry points into data exploration, and deciphering the lens encoding allowed visitors to make their first data interpretations.

In fact, we found evidence that color was not just ignored but problematic for some visitors. Over half of the visitors (24/44) who noted color at some point mapped color to an unintended, or 'incorrect' meaning (Table 6), and 12 of the 44 dyads ended their think-aloud with the wrong referent. Taking a closer look at how visitors tried to decode color, particularly because it was problematic, provided us with an opportunity to consider what resources visitors draw on when confronted with an unfamiliar encoding and what makes an encoding difficult to decipher.

Although incorrect in the sense that they were unintended by the visualization designers, upon closer inspection, these alternative mappings for color have rational explanations. First, the swirling colors superficially resemble depictions of currents. The different plankton populations changing over time appear as eddies in certain places, which can reinforce the color→current mapping. As one dyad explained: "*Because whatever this is, it's causing some kind of trail that goes across here. That is why I thought it was ocean current (Dyad01)*." This echoes a well-known tendency on the part of domain novices to focus and rely on superficial similarities in making sense of the unfamiliar [33]. Also, plankton are defined as life that drift with the current. Therefore, the movement of the different plankton types can easily signify ocean currents. And, while currents are readily associated with the ocean, visitors may be less inclined to think about microbial life when asked to think about the ocean.

Table 6. Examples of incorrect decoding comments for color. The three most frequent incorrect mappings are shown.

| Incorrect Mapping (count) | Example Visitor Quotes |
| --- | --- |
| color → an environmental variable (11) | Dyad21: No, it represents the various combinations of light, nitrogen and silica combinations. |
| | Dyad14: Oh [color is] different degrees of light |
| | Dyad6: Silica is basically the purple (note: diatoms are purple and depend on the presence of silica in the water) |
| color → current (11) | Dyad16: I guess it's [color is] meant to show currents or something |
| | Dyad31: This would be the, what? Mid-Atlantic currents. |
| | Dyad6: It kind of looks like a current. |
| color → temperature (9) | Dyad63: Is this water temperature? It must be. |
| | Dyad44: We just assume that these ones are colder water and blueish streams that this is some colder water too. |

Second, the bands of color show longitudinal differences as would temperature, with purple in the colder polar regions and green in the equatorial areas: "*This color, I think it must be temperature …Because look on the poles, it's the purple, then blue, then green (Dyad53).*" Given the common use of colors in weather maps to indicate temperature, it's not far-fetched for visitors to map color to temperature. And, temperature roughly correlates with light levels, which affect the type of plankton and their distributions.

Likewise, because environmental conditions such as nitrogen and silica correlate with plankton type, it is understandable that visitors might think that colors represent these variables instead of plankton. It is also possible that visitors used color as a shorthand for the environmental variables that determine the plankton that live in different parts of the ocean at different times of the year.

Visitors, therefore, could slip from one color decoding to another throughout their exhibit interaction as Dyad44's think-aloud illustrates:

Transcript excerpt from Dyad44

| | | |
|---|---|---|
| 1:00 | So the colors have to mean something. So they are telling us what plankton are in different parts of the ocean? | proposes color → plankton |
| 11:19 | See, it's getting entirely black here (points to the south pole) and moving up. Basically, what is it? The colors? | asks what the colors mean |
| 11:47 | So, do you think the black means no plankton are growing there at that time? | proposes color → plankton |
| 11:48 | Yes. | |
| 12:12 | So, it's getting black in summer. So, it's getting, I don't know, colder water? | |
| 15:25 | I think the blue is colder water than the green. | talks about temperature |
| 15:42 | There is a little blue here, and here is blueish and there is plenty of blue streams here off the coast | |
| 16:15 | I think it's cool that the coast of Africa has the blue around it. Namibia. I think it's cool that temperature changes right around the land. The colors. | |
| 16:31 | We just assume that these ones are colder water and bluish streams that this is some colder water too. | slips into color → temperature |
| 16:41 | It would be nice to have a key of what the colors are in temperature are, if that is in fact that they are. | |

In their work defining the NOVIS model, Lee et al. [19] observed that novices do not tend to question or revise their understanding of a visual encoding once established. To see if this was the case for *Plankton Populations*, we looked through the decoding statements to identify instances where a dyad changed their understanding of the color encoding from incorrect to correct and, more importantly, to determine what triggered visitors to revise an incorrect decoding.

We found a few[3] examples of dyads challenging and revising their prior decoding assumption, although these cases were rare. These events occurred when an erroneous decoding assumption led to an interpretation that contradicted their prior knowledge:

Dyad01: I don't think it's just ocean currents because doesn't the Pacific current come up this way and then go down the coast? It doesn't seem to be doing that.

Dyad16: I guess it's meant to show currents or something… But I'm not following-- since currents go this way (gestures along north-south axis on map).

Dyad63: It can't be [temperature] because isn't there a gulf stream that goes up here?

---

[3] Looking through their think-aloud data, we found that eight out of the 44 dyads, who talked about color at some point, questioned or challenged prior mappings and eventually arrived at the correct decoding for color.

Otherwise, inconsistencies, such as small eddies of colors when the dyad thought color mapped to temperature, were either not seen or ignored. Prior work in graph comprehension highlight the interplay between content knowledge and decoding [34]. The more a person knows about the subject matter the more easily they could decipher an encoding. However, visitors using visualizations of complex scientific dataset come with limited knowledge of the content. In fact, these visualizations are often designed to introduce a new phenomenon to its users.

**Design Implications**. It is important to provide a clear one-to-one mapping between a representation and the data represented; otherwise, visitors can become confused, sliding among possible mappings and questioning which is the correct one. Since visitors may depend on superficial appearances or more familiar uses of an encoding in determining its referent, encodings need to be carefully chosen while considering how they are typically used in other public venues. Finally, a part of helping visitors decode a visualization may depend on identifying and encoding an aspect of the complex dataset that they are familiar with, if at all possible. Visitors then have the opportunity to see inconsistencies between a suspect mapping and their prior knowledge and revise an initially erroneous decoding.

### 5.5 MERs that Constrained and Complemented Helped Decoding

*Plankton Populations* encodes the data in a complex dataset across multiple, related visual representations, or MERs. Although MERs could add to the complexity of the visualization, we found examples of different visual encodings constraining and complementing each other to help visitors' decoding efforts. Looking through visitors' think-aloud data, we found evidence of visitors using a more readily decoded visual representation to decipher a more confusing or unfamiliar encoding. For example, although a map is a familiar representation for most visitors [9], most maps are of land, and focusing on the ocean was enough of a change to confuse some (14/56) visitors: "*First it looks like this is the Earth (points to ocean), but no, this is the Earth (points to land) (Dyad65).*" The text 'No plankton on land,' which appears within a lens when it is positioned over a landmass, helped many (12/14) of these disoriented visitors read the marine map: "*[Reading] No plankton on land. Oh, this is the land. I thought it was the other way around (Dyad13).*" Although other dyads thought 'No plankton on land' was superfluous, the text in the lens, nonetheless, served an important role in helping those visitors who did not readily recognize a map of the oceans to switch perspectives.

We also found examples of visitors changing their decoding in an attempt to reconcile the referent of the plankton icons in the lenses with the colors on the map, another example of what Ainsworth describes as the constraining function served by MERs [17]. The following excerpt from Dyad13's think-aloud illustrates this constraining function at play, in which their understanding of the lens was used to limit the decoding possibilities for the map's colors.

Transcript excerpt from Dyad13

| | | |
|---|---|---|
| 01:58 | Maybe it [colors] shows what pollution does. | proposes color → pollution |
| 03:46 | I'm trying to figure out why different colors show different pictures | |
| 03:50 | Look at the picture on the purple one and look at the picture on the green one and look at the picture on the brown one (moves lens around to different colors on map) | links colors to lens icons |
| 08:15 | check over there (points to swirling colors) | |
| 10:47 | I feel like the single color only has its own, and when you get to the half and half, it starts to mix them. | |
| 10:58 | There's a little bit of blue there. So that's why you see all three here, but not here. | |

Mapping color→ plankton was sometimes helped by the static legend in the exhibit's label: "*So as I go through the purple [on the map], I see these little guys [diatoms]. And, when I go here [bluish band on map], I see those guys [blue dinoflagellate planktons] and the dots (points to the legend) (Dyad10).*" However, because the colors on the map change hues and saturations with the changing mix and concentration of different plankton types, the link between the lens icons and the colors on the map was not always apparent to visitors. Furthermore, the legend that let visitors know what the icons represented was on the perimeter of the exhibit and could remain unseen and, therefore, unhelpful in any deciphering effort. Only five dyads mentioned that the legend helped them connect the icons within the lenses to the colors on the map.

Alternatively, complementary encodings (e.g., the colors on the map and the lens icons representing plankton distribution) allowed visitors who were unable to decode the colors another way of accessing the data for exploration. As described earlier, we found that a majority of visitors' first data interpretation comments were about the location of the plankton (29/55), and that all of these dyads used the lens icons to make that interpretation. This complementary representation, therefore, was particularly useful given the trouble visitors had decoding color.

**Design Implication**. Visual encodings that are intended to constrain decoding should be placed in close proximity. This is less critical for representations designed to complement one another. Although subsequent linking between these different encodings can build a deeper understanding of the underlying data, multiple complementary ways into the data may be the more critical design criteria to ensure initial access and a quick entry into data exploration.

## 6    LIMITATIONS

This study provided only one case study of a complex process. While we hope the findings described in this paper can advance the field, future work should consider the limitations of this study. First, the findings predominantly relied on analysis of think-aloud data collected from visitors recruited to use *Plankton Populations*. Yet, not everything visitors were thinking was or could easily be verbalized. Furthermore, verbal reporting could affect and be affected by what visitors were thinking and doing, and this effect can become more pronounced when the subject is under high cognitive load [35]. This may have been the case for some of this study's visitors who, seemingly on-task, would fall silent as they struggled to make sense of a complex data visualization. Even with prompting from the data collector, sometimes visitors had little to say.

Although naturalistic observations on the total time visitors spent at the exhibit were used to check visitors' reactivity, recorded conversations of uncued visitors would have been invaluable in determining TFI and TFCI in natural behavior, especially because the cued visitors asked to think aloud may have been much more attentive and thorough in their efforts to decode and interpret the data represented. Unfortunately, the Exploratorium's acoustic environment made it difficult to capture good quality audio at *Plankton Populations* without lapel microphones. And, the need to secure informed consent for any type of audio recording would have still required some type of cuing.

This study was largely a qualitative look at the decoding process and was useful in revealing aspects that helped and hindered visitors in making sense of a complex visualization. Further work would be needed to better hone the design implications that were surfaced here. For example, to better identify an optimal number of MERs, future work may involve conducting a series of comparative studies wherein complementary encodings are removed one-by-one until a core, essential set remains.

Finally, this work took place in a hands-on, interactive museum. An analysis of this same visualization in a less interactive context (e.g., an aquarium or a visitor center at a field station) may have different results [7].

## 7    CONCLUSION

This study investigated museum visitors' decoding of a visualization of complex scientific data. Our analysis of this process provided several insights that can inform the design of the increasing number of visualizations created for museums and other informal learning settings. First, we found that visitors engaged in decoding throughout their data explorations instead of learning to decipher all the encodings all at once. We also found that visitors did not systematically decode different elements of a visualization, but often revisited encodings, mapping and remapping a representation to a referent. This suggests the need to support decoding throughout an experience, not just during initial engagement. Second, we found that while all but one dyad could make an interpretation of the data, the median time to their first data interpretation statement was 43 seconds, and 54 seconds to their first correct interpretation, while uncued visitors behaving naturally spent a median time of 47 seconds total at the *Plankton Populations* visualization. A key challenge to designing any visualization for the museum context, therefore, lies in lowering the barriers to data exploration by enabling rapid decoding.

To investigate how different design choices may have affected visitor decoding, we analyzed the data to look at the impact of our encoding choices. We found that adding a representation of a secondary data variable seemed to increase the time to first interpretation, despite efforts to downplay that variable in favor of the primary data. This suggests that when designing visualizations for museums, additional representations should be carefully considered and secondary data may need to be left out. This may be an especially difficult decision, as one of the exciting aspects of visualization is the ability to explore the connections between different datasets. In addition, encodings need to be carefully designed to avoid being easily mapped to multiple referents that seem reasonable to visitors because of their common use or superficial resemblance to familiar depictions of phenomena. Finally, we found evidence that MERs did support visitors' ability to decode. In particular, constraining MERs worked best when they were encountered or noticed simultaneously by visitors, while complementary MERs allowed visitors access to the data if one of the encodings proved to be difficult to decipher. More importantly, considering how MERs should function with one another can help inform the design of a visualization for complex datasets.

Visualizations of complex data are increasingly central to our understanding of the world, providing new insights into critical scientific topics such as climate change, genomics, and epidemiology. Museums provide a unique opportunity to engage the public with the new scientific insights and data literacy skills afforded by visualizations. By analyzing how museum visitors interpret a visualization of complex data, this paper provided new insights into how different elements aided or hindered decoding, and points to ways of better supporting the decoding process to help inform the future design of visualizations in settings of informal learning.

## ACKNOWLEDGMENTS

The authors wish to thank Lisa Sindorf, Meghan Kroning, Sophia Becker, Rachel Cushing, Leah Humphreys, Tamara Kubacki, Renae Lessing, and Martha Oropeza for their evaluation assistance. We'd also like to thank Gregory Guterman, Kevin Boyd, and Nina Fujikawa for their role in the exhibit design. This material is based upon work supported by the National Science Foundation under DRL grants 1322828 and 1323214. Any opinions, findings, and conclusions or recommendations expressed in this material are those of the authors and do not necessarily reflect the views of the National Science Foundation. Additional funding for this work was provided by the Gordon and Betty Moore Foundation.